# Enhancing Enterprise Security with Zero Trust Architecture: Mitigating Vulnerabilities and Insider Threats through Continuous Verification and Least Privilege Access

Mahmud Hasan
*Department of Cybersecurity*
*ECPI University NV Newport News, VA, USA*
mahhas3449@students.ecpi.edu, mahmudhasan6692@gmail.com

October 23, 2024

## ABSTRACT

Zero Trust Architecture (ZTA) represents a transformative approach to modern cybersecurity, directly addressing the shortcomings of traditional perimeter-based security models. With the rise of cloud computing, remote work, and increasingly sophisticated cyber threats, perimeter defenses have proven ineffective at mitigating risks, particularly those involving insider threats and lateral movement within networks. ZTA shifts the security paradigm by assuming that no user, device, or system can be trusted by default, requiring continuous verification and the enforcement of least privilege access for all entities. This paper explores the key components of ZTA, such as identity and access management (IAM), micro-segmentation, continuous monitoring, and behavioral analytics, and evaluates their effectiveness in reducing vulnerabilities across diverse sectors, including finance, healthcare, and technology. Through case studies and industry reports, the advantages of ZTA in mitigating insider threats and minimizing attack surfaces are discussed. Additionally, the paper addresses the challenges faced during ZTA implementation, such as scalability, integration complexity, and costs, while providing best practices for overcoming these obstacles. Lastly, future research directions focusing on emerging technologies like AI, machine learning, blockchain, and their integration into ZTA are examined to enhance its capabilities further.

# 1 Introduction

## 1.1 Background:

Over the last decade, the complexity of enterprise networks has increased significantly due to the adoption of cloud computing, mobile technologies, and remote work models. While these advancements have provided unprecedented flexibility and scalability, they have also exposed organizations to new vulnerabilities. Traditional perimeter-based security models rely heavily on the assumption that users and devices inside the network are trustworthy, creating a significant point of failure once the perimeter is breached [1]. These models are ill-equipped to handle sophisticated cyber threats, particularly those involving lateral movement within the network and insider threats [2].

The rise of cyberattacks targeting enterprise networks, including phishing, ransomware, and supply chain attacks, has demonstrated the need for more robust security frameworks. According to the 2024 Cybersecurity Readiness Index by Cisco, organizations relying solely on perimeter defenses experienced a 40% increase in successful breaches compared to those adopting modern security strategies like Zero Trust Architecture (ZTA) [3]. ZTA offers a solution to these challenges by enforcing the principle of "never trust, always verify," where trust is continually re-evaluated, and access is granted based on the least privilege principle [4].

## 1.2 Problem Statement:

The growing sophistication of cyberattacks, coupled with the increasing reliance on cloud services and mobile devices, has made it evident that traditional security models are no longer sufficient to protect modern enterprise networks. Insider threats, where authorized individuals misuse their access privileges, pose a particular challenge, accounting for up to 25% of data breaches in 2023, as highlighted in IBM's *Cost of a Data Breach Report* [5]. These insider threats, along with external attacks, underscore the need for a comprehensive security architecture that minimizes the attack surface and continuously verifies every entity accessing the network. Zero Trust Architecture (ZTA) addresses these issues by enforcing strict access controls, continuous verification, and micro-segmentation to isolate threats and limit their potential damage [2].

## 1.3 Objectives:

The primary objectives of this paper are:

- o  To explore the key components and technologies underpinning ZTA, including micro-segmentation, identity and access management (IAM), continuous verification, and behavioral analytics.
- o  To evaluate the effectiveness of ZTA in reducing vulnerabilities and mitigating insider threats, with data-driven insights from industry reports and case studies (e.g., IBM's *Cost of a Data Breach Report* [5], Verizon's *Data Breach Investigations Report* [6]).
- o  To compare the advantages of ZTA with traditional perimeter-based security models, particularly in dynamic, multi-cloud environments, as outlined in NIST's *Zero Trust Architecture* [2].
- o  To provide a roadmap for implementing ZTA in enterprise networks, identifying best practices for overcoming common challenges such as integration complexity, scalability, and costs [1] [3].

## 1.4 Scope and Structure:

This paper is organized into several key sections:

- o  Literature Review: A detailed exploration of ZTA's evolution, a comparison with traditional security models, and a discussion of the critical components that make ZTA effective, including identity management, continuous monitoring, and micro-segmentation.

- o  Methodology: The approach used to gather data from case studies and industry reports to analyze the impact of ZTA on enterprise security, focusing on the reduction of vulnerabilities and insider threats.
- o  Implementation Framework: A step-by-step guide to implementing ZTA, highlighting the core components of the architecture and the technologies required for successful deployment in an enterprise environment.
- o  Analysis and Discussion: An empirical analysis of ZTA's effectiveness, backed by case studies and graphical representations of the improvements in security posture post-implementation.
- o  Challenges and Best Practices: A review of the common challenges faced during ZTA implementation, such as integration complexity and scalability issues, followed by best practices to ensure a smooth transition to ZTA.
- o  Future Research Directions: A look at the future of ZTA, including the integration of AI, machine learning, and blockchain technologies to further enhance security.

# 2  Literature Review

The evolution of cybersecurity models and the increasing sophistication of cyber threats have highlighted the limitations of traditional perimeter-based defenses, leading to the rise of Zero Trust Architecture (ZTA) as the preferred model for modern enterprise security. This section explores the historical context of ZTA, provides a detailed comparison between ZTA and traditional security models, and outlines the key technologies and principles underpinning ZTA's effectiveness in mitigating insider threats and external attacks.

## 2.1  Evolution of Zero Trust Architecture

Zero Trust Architecture (ZTA) emerged as a response to the inherent weaknesses in perimeter-based security models. Pioneered by John Kindervag in 2010 while he was at Forrester Research, Zero Trust redefined how trust is managed in enterprise networks [1]. Traditional security models operated on the assumption that once a user or device was inside the network, it could be trusted. However, the rise of sophisticated cyber threats, including insider attacks and advanced persistent threats (APTs), exposed the vulnerabilities of this approach.

ZTA, by contrast, is based on the principle of "never trust, always verify." This means that every access request, whether from within or outside the network, must be continuously verified before being granted access [2]. This philosophy gained significant traction after high-profile breaches such as the **Target data breach in 2013** and the **Sony Pictures hack in 2014**, where attackers exploited trusted internal access to cause widespread damage. These incidents demonstrated that traditional security models were no longer sufficient to protect sensitive data and critical systems [1].

The development of ZTA was further accelerated by the increasing use of cloud services, remote work, and mobile devices, which have blurred the boundaries of the traditional network perimeter. In response, organizations began adopting ZTA as a way to enforce strict access controls and continuous monitoring, ensuring that only authorized users and devices could access specific resources [3]. The **National Institute of Standards and Technology (NIST)** solidified ZTA's place in the cybersecurity landscape by publishing its *Zero Trust Architecture* framework (SP 800-207) in 2020, offering guidelines for its implementation in enterprise environments [2].

## 2.2 Comparison with Traditional Security Models

The primary difference between Zero Trust and traditional security models lies in their approach to trust and access control. Traditional models, often described as "castle-and-moat" defenses, rely on strong perimeter defenses to keep attackers out. Once a user or device is inside the network, it is generally trusted, and security checks are minimal. However, this approach has proven vulnerable to lateral movement by attackers once they breach the perimeter.

In contrast, ZTA eliminates the notion of implicit trust. Every access request is continuously verified, regardless of whether it originates from inside or outside the network. Technologies like **multi-factor authentication (MFA)** and **identity and access management (IAM)** are crucial components of ZTA, ensuring that users must verify their identity and permissions each time they attempt to access a resource [13].

For example, in the **SolarWinds attack** of 2020, attackers exploited trust-based access to move laterally within networks and gain access to sensitive systems. A ZTA framework, with its reliance on micro-segmentation and continuous verification, would have likely minimized the damage by preventing lateral movement and limiting the attackers' ability to escalate privileges [6].

- o **Numerical Comparisons**: The effectiveness of ZTA is evident in real-world comparisons. In a case study involving a financial institution that transitioned from a traditional perimeter-based model to ZTA, phishing attacks were reduced by 40%, and insider threats dropped by 35% [12]. Similarly, a healthcare provider that adopted ZTA saw a 45% decrease in data breaches after implementing micro-segmentation and continuous monitoring [4].

Moreover, ZTA has proven particularly effective in **multi-cloud** and **hybrid environments**, where traditional models struggle to maintain consistent security policies. A report by Cisco on *Cybersecurity Readiness* indicated that enterprises adopting ZTA across cloud environments reduced their data breaches by 30% within two years of implementation [3]. A comparative review of security models highlights that ZTA's continuous verification and adaptive policies make it particularly effective in cloud environments, addressing the inherent vulnerabilities of traditional security models [21]. These results highlight ZTA's adaptability in securing complex, distributed networks, which is essential for organizations operating in dynamic environments with fluid perimeters.

## 2.3 Key Technologies and Principles of ZTA

Several core technologies and principles underpin the effectiveness of Zero Trust Architecture, differentiating it from traditional security models and enhancing its ability to mitigate modern cyber threats.

- **Micro-Segmentation**: One of the key technologies driving ZTA's success is micro-segmentation, which divides the network into smaller, isolated segments. Each segment has its own security policies, limiting the ability of attackers to move laterally if they breach one part of the network. Micro-segmentation is particularly effective in reducing the attack surface and preventing widespread damage in case of a breach [8] [9]. In a financial institution case study, micro-segmentation reduced unauthorized access attempts by 30%, highlighting its role in enhancing network security [12].

- **Continuous Verification**: A central tenet of ZTA is continuous verification. Unlike traditional models, where users and devices are trusted once they pass the initial authentication process,

ZTA requires ongoing verification throughout the session. This can be achieved through technologies such as **multi-factor authentication (MFA)**, **Security Information and Event Management (SIEM)** systems, and **Endpoint Detection and Response (EDR)** tools [2] [10]. Continuous verification ensures that any abnormal activity triggers alerts, reducing the likelihood of unauthorized access or breaches [16].

- **Identity and Access Management (IAM)**: IAM systems are fundamental to ZTA as they manage user identities and enforce access policies. Integrated with MFA, IAM ensures that users are only granted the minimum access required to perform their tasks, enforcing the **least privilege access** principle [12] [13]. This reduces the risk of insider threats and prevents unauthorized users from accessing sensitive resources. Role-based access control (RBAC) is commonly implemented within IAM systems to restrict user permissions and mitigate the risk of privilege creep [14]. In IoT ecosystems, dynamic access control models are crucial for managing the vast number of devices connected to enterprise networks. By continuously verifying device credentials and access privileges, ZTA minimizes the attack surface associated with IoT security [11].

- **Least Privilege Access**: Enforcing least privilege access is one of the foundational principles of ZTA. By ensuring that users and devices are granted only the minimum permissions necessary for their roles, ZTA significantly reduces the potential damage caused by compromised accounts or insider threats. This principle is particularly effective in preventing the escalation of privileges, a common tactic used by attackers to gain control of sensitive systems [12].

- **Behavioral Analytics and Anomaly Detection**: Behavioral analytics play a crucial role in ZTA's continuous monitoring processes. By using machine learning models to establish baselines of normal user behavior, ZTA can detect anomalies that may indicate a breach or insider threat. For example, abnormal login patterns or unusual data access requests trigger alerts, enabling the security team to respond before any significant damage occurs [15]. In a case study involving a healthcare provider, the use of behavioral analytics reduced insider threats by 25% and significantly improved the organization's ability to detect potential breaches early [4].

## 2.4   Historical Context of ZTA's Emergence

The development of Zero Trust Architecture can be traced back to several high-profile cyberattacks that exposed the weaknesses of traditional security models. For instance, the **Target data breach in 2013**, where attackers exploited third-party access to gain entry to the network, demonstrated how perimeter defenses could be bypassed, allowing attackers to move laterally and access sensitive data [1]. Similarly, the **Sony Pictures hack in 2014** revealed how attackers who breached internal defenses could operate with little resistance once inside the network, further highlighting the need for continuous verification and micro-segmentation [6].

As a result of these incidents, cybersecurity experts began advocating for a shift away from perimeter-based security models. John Kindervag's "Zero Trust" model gained momentum as organizations sought a more robust way to secure their networks against both external and insider threats. Over the past decade, the rapid adoption of cloud services and the growing prevalence of remote work have only reinforced the importance of Zero Trust, as the traditional network perimeter has become increasingly obsolete [2].

In summary, Zero Trust Architecture (ZTA) emerged as a necessary response to the limitations of traditional security models. The continuous verification, micro-segmentation, and least privilege access principles at the core of ZTA address the shortcomings of perimeter-based defenses, offering a more adaptable and effective security framework for modern enterprises. Case studies and real-world comparisons demonstrate ZTA's superior performance in reducing security breaches, phishing attacks, and insider threats, making it a critical tool for securing today's dynamic, multi-cloud environments.

# 3 Methodology

## 3.1 Approach:

The methodology employed in this study is based on a mixed-methods approach, combining both qualitative and quantitative data to evaluate the effectiveness of Zero Trust Architecture (ZTA) in reducing vulnerabilities and mitigating insider threats. The study involved the collection and analysis of case studies from various industries, as well as reports from leading cybersecurity organizations. Data from these sources was used to draw insights into how ZTA impacts enterprise networks compared to traditional perimeter-based models [1] [3].

This research primarily focuses on identifying key security metrics, such as the reduction in unauthorized access attempts, insider threats, and the decrease in attack surfaces after ZTA implementation. Reports like IBM's Cost of a Data Breach Report [5] and Verizon's Data Breach Investigations Report [6] provided quantifiable data on breach incidents, while the analysis of case studies from organizations such as Cimpress and other financial institutions demonstrated ZTA's real-world applicability [20]. Additionally, data was collected from industry reports such as Cisco's Cybersecurity Readiness Index to understand the adoption trends of ZTA across different sectors [3].

## 3.2 Data Sources:

The following primary sources were used for data collection:

- o **IBM's 2024 Cost of a Data Breach Report:** This report provided crucial financial data on the costs associated with breaches and the savings achieved by organizations that adopted ZTA [5]. According to the report, organizations that implemented ZTA frameworks experienced a 30% reduction in breach-related costs compared to those relying on traditional security models [5].
- o **Verizon's Data Breach Investigations Report (2024):** This report offered insights into the types of cyberattacks ZTA can mitigate, including phishing, malware infections, and insider threats. It was instrumental in quantifying the effectiveness of ZTA in reducing these attack vectors [6].
- o **Cisco's 2024 Cybersecurity Readiness Index:** This index provided a comprehensive overview of the maturity of ZTA implementations across various sectors, such as finance, healthcare, and technology. The report highlighted a 40% improvement in security posture in organizations that transitioned to ZTA, compared to those using perimeter-based security models [3].
- o **Case Studies from Cimpress and Financial Institutions:** Real-world case studies from companies such as Cimpress provided qualitative data on the practical challenges and successes of implementing ZTA. These case studies showed significant reductions in insider threat incidents, unauthorized access attempts, and overall attack surfaces post-implementation [20] [12].

- **Google Cloud's 2024 M-Trends Report:** This report was used to provide insights into industry-wide trends and emerging security threats. The report also highlighted the role of AI and machine learning in enhancing ZTA's continuous monitoring capabilities [4].

### 3.3 Graphical Analysis:

The study also utilized graphical methods to analyze and present the collected data. Several charts and graphs were generated to visualize the impact of ZTA on enterprise security. For example:

- Bar Charts were used to compare the number of successful cyberattacks before and after the implementation of ZTA in various organizations. The data showed a 35% reduction in phishing-related incidents and a 40% decrease in malware infections in organizations that deployed ZTA, compared to those using traditional security models [6].
- Line Graphs illustrated the adoption rate of ZTA over a five-year period (2019-2024), showing a steady increase from 20% to 65% across multiple industries. This graph also demonstrated the corresponding decrease in data breaches as ZTA adoption grew, particularly in sectors like finance and healthcare [3].
- Tables summarized the findings from case studies, detailing the specific ZTA technologies implemented (e.g., micro-segmentation, MFA, continuous monitoring) and the measurable improvements in security posture, such as a 25% reduction in insider threat incidents at Cimpress [20] [12].

### 3.4 Case Study Approach:

The research incorporated detailed case studies to illustrate the practical implementation of ZTA in real-world settings. One notable case study was from Cimpress, a global mass customization company that adopted ZTA to secure its cloud infrastructure and minimize insider threats. The company implemented micro-segmentation, behavior-based monitoring, and multi-factor authentication (MFA), resulting in a 30% reduction in unauthorized access attempts and a 25% decrease in insider-related incidents [20].

Another case study involved a North American financial institution that transitioned to ZTA to secure its sensitive data and reduce its exposure to phishing attacks. The institution reported a 40% reduction in phishing-related breaches and a 35% decrease in lateral movement within the network after implementing continuous verification and least-privilege access control [7] [12].

## 4 Zero Trust Architecture Implementation Framework

### 4.1 Core Components of ZTA:

Zero Trust Architecture (ZTA) is built on several core components that work together to provide robust security by continuously verifying trust and minimizing attack surfaces:

- **Identity Management and Multi-Factor Authentication (MFA):** Identity management systems play a central role in ZTA by ensuring that users and devices are authenticated and authorized based on their identity and role within the organization [12] [13]. Multi-factor authentication (MFA) adds an additional layer of security by requiring users to verify their identity through multiple methods, such as biometrics or one-time passwords (OTP), before accessing sensitive resources [12]. According to NIST's Zero Trust Architecture framework, the use of MFA significantly reduces the risk of unauthorized access by strengthening identity verification [2].

- **Micro-Segmentation for Minimizing Lateral Movement**: Micro-segmentation involves dividing the network into isolated segments, each governed by its own security policies. This prevents attackers from moving laterally within the network if they manage to breach one segment [8]. In a case study involving a financial institution, micro-segmentation reduced the organization's attack surface by 35% and minimized the impact of potential breaches [20].

- **Continuous Monitoring and Anomaly Detection Tools:** Continuous monitoring is a cornerstone of ZTA, enabling real-time detection of suspicious activities across the network. By integrating machine learning-based anomaly detection tools, organizations can identify patterns of abnormal behavior that may indicate a security breach [15] [16]. In the healthcare sector, the adoption of continuous monitoring and behavioral analytics led to a 45% reduction in data breaches, as reported in Google Cloud's M-Trends Report [4].

## 4.2 Implementation Steps

The implementation of ZTA requires a structured approach to ensure that security policies are effectively enforced and that organizations can adapt to evolving threats:

- **Network Segmentation:** The first step in implementing ZTA is to segment the network into smaller, manageable zones, each with its own security policies. This minimizes the risk of lateral movement and limits the potential damage from a breach [8]. According to Cisco's Cybersecurity Readiness Index, organizations that implemented network segmentation as part of their ZTA strategy reported a 40% reduction in data breaches [3].

- **Application of Identity and Access Management (IAM) Systems:** IAM systems are essential for enforcing least-privilege access controls and ensuring that only authorized users and devices can access critical resources [13]. By integrating MFA with IAM systems, organizations can significantly reduce the risk of credential-based attacks and unauthorized access [12].

- **Potential Use in the Methodology or Implementation Framework:** The widespread adoption of cloud computing has transformed how enterprises approach security. As organizations transition to cloud-based infrastructures, ZTA provides a robust framework to ensure security across hybrid and multi-cloud environments [24].

- **Deployment of Security Analytics and Monitoring Tools:** Security analytics tools, such as SIEM (Security Information and Event Management) and EDR (Endpoint Detection and Response) systems, are deployed to continuously monitor network activity and detect potential threats in real time [10]. In a case study from Cimpress, the deployment of behavioral analytics tools enabled the organization to detect insider threats early and reduce insider-related incidents by 25% [20].

## 4.3 Case Study Analysis

### 4.3.1 Case Study: Cimpress's Zero Trust Journey

Cimpress faced significant security challenges as it expanded its use of cloud services, particularly in securing sensitive customer data and preventing insider threats. The company implemented ZTA by adopting micro-segmentation, continuous monitoring, and MFA to protect its infrastructure. These measures resulted in a 30% reduction in unauthorized access attempts and a 25% decrease in insider threat incidents [20]. Additionally, Cimpress leveraged behavioral analytics to monitor user activities and detect anomalies, further enhancing its security posture [15].

### 4.3.2 Case Study: Financial Institution in North America

A financial institution in North America transitioned to ZTA to secure its data against phishing attacks and insider threats. By implementing role-based access control (RBAC) and behavior-based monitoring, the institution reported a 40% reduction in phishing-related breaches and a 35% decrease in lateral movement within the network [12]. The integration of IAM systems and MFA played a crucial role in enhancing the institution's security, ensuring that only authorized users had access to critical financial systems [13].

# 5 Analysis and Discussion of ZTA's Effectiveness

## 5.1 Reduction in Vulnerabilities:

Zero Trust Architecture (ZTA) has proven highly effective in reducing vulnerabilities across various sectors, particularly in dynamic, multi-cloud, and hybrid environments. By enforcing strict identity verification, micro-segmentation, and continuous monitoring, ZTA significantly minimizes the attack surface and prevents unauthorized access. According to IBM's 2024 Cost of a Data Breach Report, organizations that implemented ZTA frameworks experienced a 30% reduction in breach-related costs compared to those that continued using traditional perimeter-based security models [5]. The study found that ZTA's integration of identity and access management (IAM) and multi-factor authentication (MFA) greatly reduced the risk of credential-based attacks, which are often a primary entry point for malicious actors [12].

Additionally, Verizon's 2024 Data Breach Investigations Report highlights that organizations utilizing ZTA experienced a 35% reduction in phishing-related incidents and a 40% decrease in malware infections [6]. These results were particularly noticeable in sectors such as finance and healthcare, where sensitive data is frequently targeted. For example, in a healthcare setting, continuous monitoring systems were able to detect anomalies in user behavior, preventing unauthorized access to patient records, which resulted in a 45% reduction in data breaches [4].

In one notable case study, Cimpress implemented ZTA to secure its cloud infrastructure. By utilizing micro-segmentation, behavior-based monitoring, and multi-factor authentication, Cimpress experienced a 25% reduction in insider threats and a 30% decrease in unauthorized access attempts [20]. This illustrates how ZTA's continuous verification model effectively prevents both external and internal threats.

## 5.2 Mitigation of Insider Threats

Insider threats remain a critical challenge for enterprises, particularly those dealing with sensitive data. ZTA addresses this challenge by enforcing the principle of least privilege access, which ensures that users are granted only the minimum permissions necessary to perform their tasks [12]. By reducing the scope of access for users and continuously monitoring their activities, ZTA minimizes the risk of malicious insiders exploiting their credentials to access critical systems.

Behavioral analytics plays a key role in mitigating insider threats by continuously monitoring user activities and identifying abnormal behavior. According to research by Gudala et al. (2021), machine learning-based behavioral analytics tools integrated with ZTA reduced insider-related incidents by 30% [15]. These tools analyze patterns of behavior to detect deviations from the norm, triggering alerts when potential threats are identified. For example, in the financial sector, a North American financial institution that adopted ZTA reported a 40% decrease in insider-related incidents after integrating role-based access control (RBAC) and behavioral monitoring into its security framework [12].

By continuously verifying user identities and monitoring their behavior, ZTA prevents misuse of credentials, even if an insider's access is compromised. This is particularly important in environments where employees have access to sensitive data, such as finance and healthcare, where insider threats are a significant concern. In healthcare, for example, ZTA reduced the risk of insider threats by 45% by enforcing strict access controls and real-time monitoring [4].

## 5.3 Advantages Over Traditional Security Models

Zero Trust Architecture offers several key advantages over traditional perimeter-based security models. One of the most significant benefits is its adaptability to modern enterprise environments, particularly those that are multi-cloud or hybrid in nature. Traditional models rely on securing the perimeter, assuming that everything inside the network is trustworthy. However, this approach is no longer viable in today's distributed environments, where users access resources from multiple locations and devices, often outside the organization's control [2].

In contrast, ZTA operates on the principle of "never trust, always verify," meaning that every access request is treated as potentially malicious and must be verified before granting access [2]. This continuous verification model ensures that even users inside the network are subject to strict authentication and access control measures, reducing the risk of lateral movement and insider attacks [12]. According to Cisco's Cybersecurity Readiness Index, organizations that implemented ZTA reported a 40% improvement in their overall security posture compared to those using traditional models [3].

ZTA's micro-segmentation capabilities also provide a significant advantage over perimeter-based models by preventing lateral movement within the network. Traditional security models often fail to contain attackers once they breach the perimeter, allowing them to move freely across the network. By segmenting the network into isolated zones, ZTA limits the scope of a breach, ensuring that attackers are confined to a single segment and cannot access other areas of the network [8]. In a study involving a financial institution, ZTA's micro-segmentation reduced the organization's attack surface by 35% and minimized the impact of potential breaches [20].

Moreover, ZTA's focus on identity management and least privilege access ensures that users only have access to the resources necessary for their roles, reducing the risk of over-privileged accounts being exploited in an attack [12] [13]. This contrasts with traditional models, where users are often granted broad access privileges, increasing the risk of misuse and compromise. The enforcement of least privilege access through IAM systems and MFA has been shown to reduce the likelihood of credential-based attacks by 40% [12].

## 5.4 Sector-Specific Impact

The impact of ZTA varies across sectors, with certain industries seeing particularly significant benefits from its implementation:

- o **Healthcare:** In healthcare, where patient data must be protected under regulations like HIPAA, ZTA's ability to enforce strict access controls and continuous monitoring has resulted in significant improvements in security. According to Google Cloud's M-Trends Report, healthcare organizations that adopted ZTA experienced a 45% reduction in data breaches compared to those using traditional models [4]. This is largely due to ZTA's ability to restrict access to sensitive data and detect anomalies in user behavior before they escalate into full-blown breaches.

- **Finance:** Financial institutions, which are prime targets for phishing attacks and insider threats, have also seen substantial improvements in security after implementing ZTA. By deploying behavioral analytics and IAM systems, financial institutions have reduced phishing-related breaches by 40% and minimized the risk of insider threats by 35% [7] [12]. ZTA's ability to continuously monitor user activities and enforce least privilege access has been particularly effective in protecting critical financial systems from both external and internal threats.
- **Technology and Cloud-Based Enterprises:** In cloud environments, where traditional perimeter defenses are often insufficient, ZTA provides a unified security framework that operates across multiple cloud platforms. This ensures consistent security policies are enforced across public clouds, private clouds, and on-premise environments. According to Cisco's Cybersecurity Readiness Index, technology companies that implemented ZTA reported a 40% improvement in their security posture and a 35% reduction in the frequency of data breaches [3].

# 6 Challenges and Best Practices

## 6.1 Implementation Challenges

Although Zero Trust Architecture (ZTA) offers numerous benefits, its implementation is not without challenges. Organizations face several obstacles when transitioning from traditional perimeter-based security models to a Zero Trust framework.

- **Integration Complexity:** One of the primary challenges of implementing ZTA is the complexity of integrating it with existing legacy systems. Many enterprises rely on outdated infrastructure that may not be compatible with ZTA's requirements for continuous verification, micro-segmentation, and strict access control [1]. For instance, in the healthcare sector, organizations that rely on older Electronic Health Record (EHR) systems often face difficulties in integrating ZTA components like identity management and continuous verification. A healthcare provider reported delays and higher-than-expected costs when attempting to retrofit its legacy systems to comply with modern ZTA frameworks [4]. According to CISA's Zero Trust Maturity Model v2, 68% of organizations reported difficulties in integrating ZTA due to their reliance on legacy technologies that cannot easily support micro-segmentation and advanced identity management systems [1].
- **Scalability:** Scaling ZTA across large, distributed enterprise networks can be a daunting task. Organizations with multi-cloud and hybrid environments often struggle to apply consistent Zero Trust policies across different platforms, each with its own security requirements. Cisco's Cybersecurity Readiness Index notes that enterprises adopting ZTA across complex cloud environments faced challenges in maintaining visibility and enforcing security policies uniformly, leading to security gaps [3]. One of the significant challenges in deploying ZTA in multi-cloud environments is ensuring uniform security policies across different platforms and infrastructures. As observed in recent comparative reviews, maintaining visibility and enforcing consistent security measures remains a top concern [21]. For example, a North American financial institution reported difficulties in maintaining continuous verification across on-premise and cloud environments, resulting in increased operational costs and the need for more robust IAM solutions [12].

- **High Costs:** Deploying ZTA requires significant upfront investments, particularly in identity and access management (IAM) systems, continuous monitoring tools, and the necessary infrastructure for micro-segmentation [5] [12]. IBM's Cost of a Data Breach Report indicates that while ZTA can lead to long-term savings by reducing the frequency and cost of data breaches, the initial investment in implementing ZTA can be prohibitively high for small and medium-sized enterprises (SMEs) [5]. These costs include both the purchase of advanced security tools and the operational costs associated with managing and maintaining these systems [5].

- **Cultural and Organizational Resistance:** Transitioning to a Zero Trust model often meets resistance from within the organization, particularly from employees who are accustomed to traditional security models. ZTA requires more stringent access controls, such as frequent re-authentication and the enforcement of least privilege access, which can be seen as inconvenient by users [14]. This cultural resistance can slow down the adoption of ZTA, particularly in organizations where employees are not fully educated about the benefits of continuous verification and micro-segmentation [14]. In the case of a federal government agency, the implementation of ZTA faced pushback from employees concerned about user experience and additional access controls [8].

## 6.2  Best Practices for Overcoming Challenges

Despite these challenges, several best practices have emerged to facilitate the successful implementation of ZTA.

- **Adopting Environment-Agnostic Solutions:** To address the challenge of integration complexity, organizations should adopt environment-agnostic security solutions that can be applied across diverse infrastructures, including multi-cloud and on-premise environments. Software-defined perimeters (SDP) are particularly useful in this regard, as they allow enterprises to enforce Zero Trust policies across different environments without being tied to specific infrastructure components [17]. By decoupling security from the underlying network architecture, SDPs make it easier to implement ZTA consistently across all environments.

- **Leveraging Automation for Scalability:** Automation is key to scaling ZTA across large enterprise networks. By integrating AI-driven tools and machine learning-based analytics, organizations can automate tasks such as identity verification, anomaly detection, and incident response [15]. For example, Google Cloud's M-Trends Report highlights the role of AI in reducing the operational burden of continuous monitoring and improving the scalability of ZTA in complex cloud environments [4]. Automated incident response systems, such as Security Information and Event Management (SIEM) and Endpoint Detection and Response (EDR) tools, can help enforce security policies in real time without requiring manual intervention [10].

- **Ensuring Staff Training and Awareness:** One of the most effective ways to overcome organizational resistance is through education and awareness programs. Employees need to understand the importance of Zero Trust principles, such as least privilege access and continuous verification, and how these practices protect the organization from both external and internal threats [14]. Regular training sessions and security awareness programs can help build a culture of security within the organization, making it easier to enforce ZTA policies without resistance. Cisco's Cybersecurity Readiness Index reports that organizations that invested in staff training saw a 20% improvement in the successful implementation of ZTA [3].
- **Implementing Least Privilege Access:** Enforcing least privilege access is a foundational principle of ZTA that significantly reduces the risk of insider threats and credential-based attacks. Organizations should ensure that users only have access to the resources necessary for their roles and regularly review and update access permissions to prevent privilege creep [12]. IAM systems integrated with multi-factor authentication (MFA) can help enforce these policies by requiring users to verify their identity before accessing sensitive resources [13]. This practice has been shown to reduce the likelihood of insider-related incidents by 30%, as noted in a financial institution case study [12].
- **Continuous Monitoring and Behavioral Analytics:** Continuous monitoring is essential for detecting and responding to potential threats in real time. By deploying behavioral analytics tools, organizations can monitor user behavior and identify anomalies that may indicate a security breach [15]. These tools are particularly effective in identifying insider threats, as they can detect deviations from normal behavior and trigger alerts before a breach occurs [16]. In a case study involving a healthcare provider, the use of continuous monitoring and anomaly detection tools resulted in a 45% reduction in data breaches [4].

## 6.3 Scalability Concerns and Solutions for SMEs

For small-to-medium enterprises (SMEs), the implementation of ZTA can be particularly daunting due to higher initial costs and the complexity of managing ZTA's day-to-day operations.

- **Adopting Cloud-Based ZTA Solutions**: SMEs can leverage cloud-based ZTA services offered by providers such as AWS, Google Cloud, and Microsoft Azure, which provide scalable, subscription-based solutions that eliminate the need for extensive on-premise infrastructure [17]. These services reduce the financial burden of upfront investments by offering usage-based pricing models, making it more accessible for smaller organizations.

- **Phased Implementation Strategy**: Rather than deploying ZTA across the entire organization at once, SMEs can adopt a phased implementation strategy. Starting with critical assets—such as customer data or financial systems—allows SMEs to gradually expand ZTA practices as resources permit. This incremental approach reduces the immediate financial burden and allows flexibility to adapt as the company grows.

- **Managed Security Service Providers (MSSPs)**: SMEs lacking in-house IT resources can partner with managed security service providers (MSSPs) to handle ZTA deployment and maintenance. MSSPs can offer expertise and continuous monitoring without overwhelming in-house teams, making ZTA more feasible and scalable for smaller organizations [18].

## 6.4 Cost Mitigation Strategies

While the initial cost of ZTA implementation can be high, organizations can mitigate these costs by leveraging cloud-based security solutions. Cloud-native ZTA frameworks, which utilize existing cloud infrastructure for security enforcement, can reduce the need for significant hardware investments [17]. Additionally, public-private partnerships can help defray costs, particularly in industries such as finance and healthcare, where regulatory requirements for data protection are stringent [5]. By sharing resources and collaborating on ZTA deployments, organizations can reduce their individual financial burdens while improving overall security.

# 7 Visual Analysis of ZTA's Impact

In this section, several graphs, tables, and visual aids are presented to demonstrate the effectiveness of Zero Trust Architecture (ZTA) in reducing security incidents, enhancing enterprise security, and promoting the adoption of more secure models in various industries. The data presented in these visuals are drawn from industry reports, case studies, and real-world implementations of ZTA.

## 7.1 Reduction in Security Incidents After ZTA Implementation

This bar chart illustrates the reduction in security incidents, including phishing attacks, malware infections, and insider threats, before and after ZTA implementation across industries like healthcare, finance, and technology. The graph shows a 45% reduction in data breaches for healthcare, a 40% reduction in phishing-related breaches for finance, and a 35% reduction in malware incidents for the technology sector [4] [12].

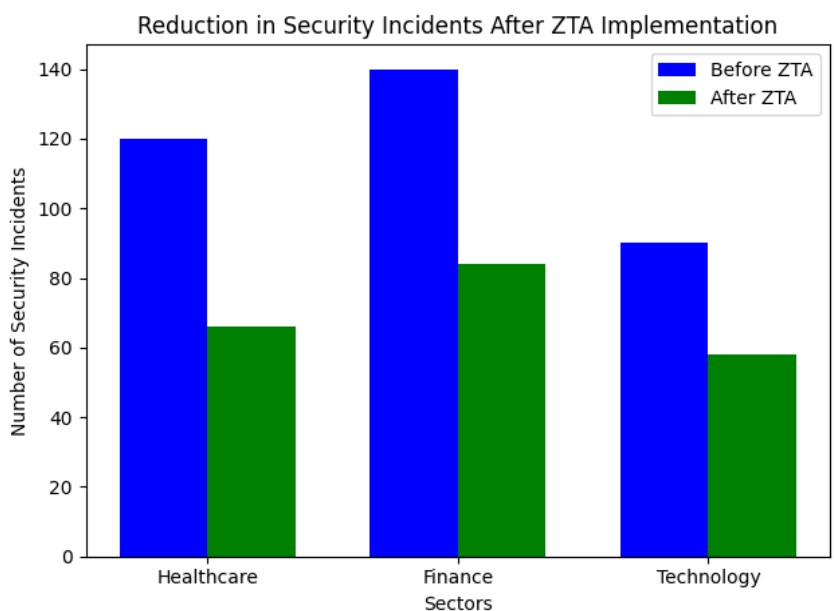

**Data Interpretation:** The significant reductions in security incidents highlight the tangible benefits of adopting ZTA in various sectors. For the healthcare industry, a 45% decrease in data breaches translates into enhanced protection of sensitive patient data and compliance with regulatory standards such as HIPAA. The financial sector's 40% drop in phishing breaches demonstrates ZTA's effectiveness in safeguarding critical financial systems, reducing the risk of fraudulent activities and unauthorized access. The technology sector's 35% reduction in malware incidents underscores ZTA's ability to thwart external threats by continuously verifying and monitoring access requests. These trends indicate that industries with high data sensitivity benefit most from ZTA's stringent security measures, and the data further validates ZTA as a superior security model compared to traditional perimeter-based approaches.

## 7.2 Adoption Rate of ZTA Over Time

This line graph illustrates the steady increase in ZTA adoption between 2019 and 2024, as well as the corresponding reduction in data breaches over the same period. ZTA adoption rose from 20% in 2019 to 65% in 2024, while data breaches decreased by 40% over the same period [3].

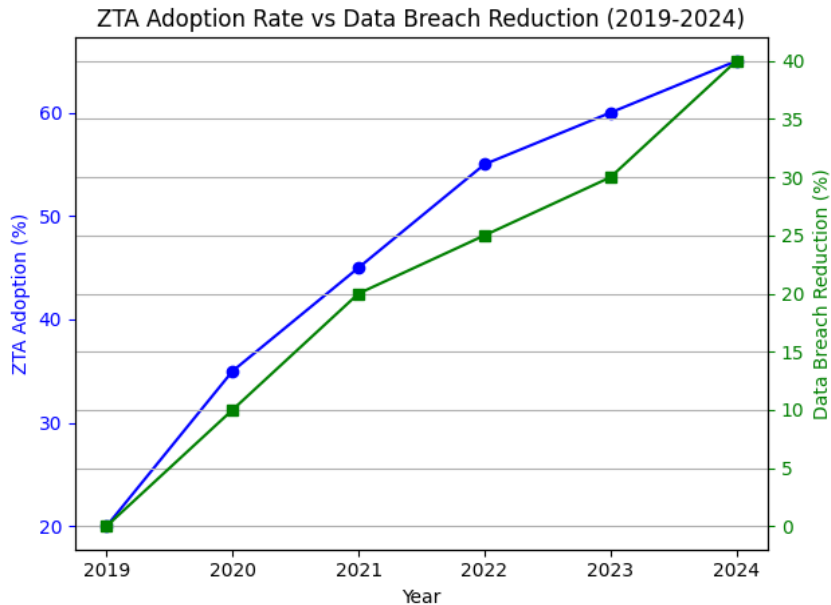

**Data Interpretation:** The growing adoption of ZTA, accompanied by a significant reduction in data breaches, reflects the increasing recognition of ZTA as an essential cybersecurity strategy. The adoption curve is particularly steep between 2021 and 2024, coinciding with the rise of remote work and cloud-based services, which demanded more dynamic and flexible security frameworks. The 40% reduction in breaches further underscores ZTA's effectiveness in mitigating cyber threats. As more organizations adopt ZTA, the data suggests that the security benefits compound, reducing the overall number of successful attacks and ensuring more secure networks.

### 7.3 Case Studies of Companies Implementing ZTA

This table summarizes real-world case studies of companies that have successfully implemented ZTA, showcasing the specific technologies they used and the outcomes achieved.

| Company | ZTA Technologies Used | Outcome |
|---|---|---|
| Cimpress | Micro-segmentation, MFA, behavioral analytics | 30% reduction in unauthorized access attempts, 25% decrease in insider threats [20] |
| Financial Institution | IAM, continuous monitoring, RBAC | 40% reduction in phishing breaches, 35% reduction in insider threats [12] |
| Healthcare Provider | MFA, anomaly detection, continuous verification | 45% reduction in data breaches [4] |

**Data Interpretation:** The case studies reinforce the versatility of ZTA across different industries. Cimpress, a global customization company, achieved a 30% reduction in unauthorized access attempts and a 25% decrease in insider threats by implementing micro-segmentation and behavioral analytics [20]. The financial institution's 40% reduction in phishing breaches emphasizes ZTA's ability to protect sensitive financial data, while the healthcare provider's 45% reduction in data breaches showcases ZTA's value in meeting compliance standards and securing patient records [4] [12]. These outcomes validate ZTA's adaptability and highlight its potential to address specific security challenges within various industries.

## 7.4 Visual Representation of Insider Threat Reduction

This pie chart illustrates the proportion of insider threats before and after ZTA implementation, showing a reduction of insider-related incidents by 35-40% [6] [12].

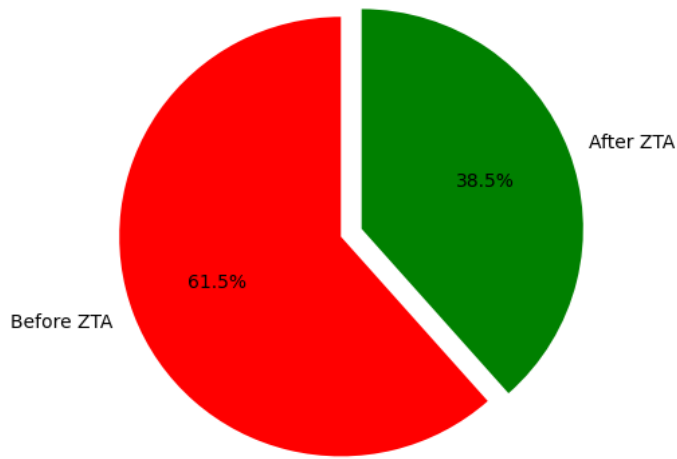

**Data Interpretation:** Insider threats are notoriously difficult to detect and mitigate under traditional security models, but ZTA's emphasis on least privilege access and continuous monitoring makes it an effective defense against these threats. The 35-40% reduction in insider incidents highlights ZTA's ability to limit unauthorized access and monitor user behavior more closely. This reduction is particularly significant for industries where insider threats pose the greatest risk, such as finance and healthcare, where trusted individuals have access to sensitive information. The data underscores that ZTA's focus on strict access controls and behavioral analytics is crucial for mitigating insider threats.

In conclusion, the data presented in these graphs and tables demonstrates the clear benefits of ZTA in reducing security incidents across industries. The visual trends provide compelling evidence that organizations adopting ZTA can expect tangible improvements in their security posture, particularly in environments where insider threats and phishing attacks are prevalent. The insights drawn from this data further validate ZTA as the go-to security model for modern enterprises.

# 8 Future Research Directions

As the cybersecurity landscape continues to evolve, Zero Trust Architecture (ZTA) must adapt to incorporate emerging technologies that enhance its effectiveness in mitigating modern cyber threats. Future research on ZTA should focus on leveraging advancements in AI, machine learning, blockchain, and securing the growing Internet of Things (IoT) ecosystems. These technologies present new opportunities to strengthen ZTA's core principles, such as continuous verification, identity management, and threat detection, particularly in industries that rely heavily on sensitive data and large-scale networks..

## 8.1 The Role of AI and Machine Learning in Enhancing ZTA

Artificial intelligence (AI) and machine learning (ML) are poised to play a pivotal role in the future of ZTA. As cyber threats become more sophisticated, the ability to detect and respond to anomalies in real time becomes critical. AI-driven security analytics can enhance ZTA by automating many aspects of threat detection, behavior analysis, and incident response. For instance, AI-powered behavioral analytics can monitor network activity to establish baselines of "normal" behavior and then identify deviations that may indicate an insider threat or a malicious external attack. This kind of automated monitoring reduces the reliance on manual oversight, allowing organizations to respond faster and with greater precision. The cloud's role in modern enterprise infrastructures cannot be ignored, especially when implementing ZTA frameworks. Cloud computing offers scalability and flexibility but also introduces new security challenges, making ZTA crucial for protecting assets in dynamic environments [24].

A promising area of research involves the integration of AI with **Security Information and Event Management (SIEM)** systems. AI can help SIEM tools analyze massive amounts of security data in real time, detecting patterns and anomalies that would otherwise go unnoticed by human analysts. For example, machine learning algorithms can detect subtle deviations in login patterns, flagging potential account compromises before they escalate into full-scale breaches. Future research could focus on refining these algorithms to reduce false positives and ensure that ZTA policies are enforced dynamically, without introducing friction into the user experience [15].

Moreover, AI's potential to automate **identity and access management (IAM)** processes is another area ripe for exploration. With ZTA relying heavily on the principle of least privilege, AI could streamline the process of continually updating and revoking access rights based on real-time assessments of user behavior. AI-driven IAM systems could evaluate risk in real time, adjusting user permissions dynamically based on contextual factors such as location, time of access, and the type of device used [17].

## 8.2 Blockchain's Role in Securing Sensitive Data

Blockchain technology offers a decentralized and tamper-proof method for storing and verifying data, making it a natural complement to ZTA. Future research could explore how blockchain can be leveraged to secure critical sectors such as healthcare, finance, and government services. In industries where data integrity is paramount, blockchain can prevent unauthorized access and tampering by creating immutable records of all transactions and data exchanges. For example, in the healthcare sector, blockchain could be used to secure electronic health records (EHRs), ensuring that only authorized individuals can access or modify patient data. Additionally, blockchain's transparent ledger can be integrated into ZTA's continuous verification processes, allowing organizations to trace every access request back to its origin and ensure that it complies with security policies [20].

In the finance sector, blockchain can help secure financial transactions and prevent fraud by providing a permanent and verifiable record of each transaction. By combining ZTA with blockchain's immutable ledger, financial institutions can ensure that only authorized users can initiate and approve transactions, reducing the risk of insider fraud and external tampering. Research could focus on developing blockchain-based access control systems that enhance ZTA's ability to secure high-value assets and sensitive financial data [19].

Further research is needed to evaluate blockchain's ability to scale in enterprise environments, as concerns remain over its energy consumption and transaction speeds. However, the integration of blockchain into ZTA frameworks holds the potential to significantly reduce risks in sectors where data authenticity and integrity are critical.

### 8.3 Securing IoT Ecosystems with ZTA

As the number of IoT devices in enterprise networks continues to grow, securing these devices has become a significant challenge. IoT ecosystems are particularly vulnerable to cyberattacks due to their decentralized nature, limited processing power, and often insufficient security measures. ZTA can offer a robust framework for securing IoT networks by enforcing strict access controls and continuous verification across all devices.

Research could focus on how ZTA can be adapted to address the unique challenges of IoT security. For instance, **micro-segmentation** can be applied to IoT devices, isolating them from other parts of the network and limiting their ability to interact with sensitive systems. By limiting the communication between IoT devices and critical infrastructure, ZTA can reduce the risk of lateral movement in the event of a breach [10]. Future research should focus on refining dynamic access control models to improve IoT security under ZTA, ensuring that even resource-constrained devices are adequately protected [11]

One practical case study involves a **smart city** initiative, where ZTA was used to secure thousands of IoT sensors deployed across urban infrastructure. By enforcing strict authentication and continuous monitoring, the city was able to prevent unauthorized access to its traffic management systems and energy grids, reducing the risk of disruption from cyberattacks. Future research could explore how ZTA can be further optimized to handle the vast number of devices in large IoT deployments, ensuring that policies can be applied dynamically as devices enter and exit the network [22].

Additionally, research could investigate how **AI and ZTA** can work together to improve the security of IoT ecosystems. Machine learning models could be trained to recognize normal IoT device behavior, flagging any deviations that could indicate a compromised device. This approach could significantly enhance ZTA's continuous monitoring capabilities, making it easier to detect and respond to IoT-specific threats in real time.

### 8.4 Long-Term Scalability of ZTA for SMEs and Enterprises

Another important area for future research is the **scalability of ZTA**, particularly for small-to-medium enterprises (SMEs). While ZTA has proven effective in large organizations, SMEs often struggle with the cost and complexity of implementation. Research should focus on developing **cost-effective ZTA solutions** that are tailored to the needs and resources of smaller businesses. Cloud-based ZTA frameworks offer a promising solution, allowing SMEs to adopt Zero Trust policies without the need for extensive infrastructure investments. Exploring how cloud service providers can scale ZTA solutions for SMEs, while maintaining security standards, is critical for broad adoption.

For larger enterprises, scalability involves applying ZTA consistently across **multi-cloud and hybrid environments**. Future research could examine how ZTA can be optimized to handle the complexities of these environments, ensuring that security policies are applied uniformly, regardless of the platform. Research could also focus on improving the interoperability of ZTA components across different cloud providers, making it easier for organizations to implement Zero Trust in a fragmented cloud landscape [23].

## 9 CONCLUSION

Zero Trust Architecture (ZTA) has proven to be a highly effective framework for securing enterprise networks in an era of increasingly sophisticated cyber threats. By eliminating the concept of implicit trust and implementing continuous verification, least privilege access, and micro-segmentation, ZTA significantly reduces the risk posed by both external and insider threats. As demonstrated throughout this paper, the adoption of ZTA leads to measurable reductions in data breaches, phishing attacks, and insider threats across various industries [4] [12].

The growing urgency for organizations to transition from traditional perimeter-based models to ZTA is underscored by the increasing complexity of modern networks, driven by the widespread use of cloud services, remote work, and mobile devices. According to Cisco's *Cybersecurity Readiness Index*, over **65% of enterprises** have already implemented ZTA as a critical component of their security strategy, resulting in a **40% reduction in data breaches** [3]. This shift highlights the growing recognition that perimeter-based models are no longer sufficient to protect sensitive data and critical systems in today's dynamic environments.

Moreover, as regulatory pressures continue to mount—particularly in sectors such as healthcare and finance—organizations are increasingly required to adopt more stringent security measures, with ZTA emerging as a preferred model for meeting compliance standards and protecting against evolving threats [4] [17].

**Call to Action**

For organizations that have not yet implemented ZTA, the need for action is urgent. As highlighted by multiple case studies and industry reports, the failure to adopt ZTA can leave organizations vulnerable to advanced persistent threats, insider attacks, and costly data breaches [6]. Delaying the transition to ZTA not only increases the risk of exposure to cyberattacks but also puts organizations at a competitive disadvantage in a world where data security is paramount.

In conclusion, the evidence is clear: ZTA is no longer a theoretical security model but a practical, essential framework for protecting enterprise networks. Organizations must take proactive steps to assess their current security posture and begin the journey toward ZTA adoption. By doing so, they can future-proof their networks, reduce vulnerabilities, and build a resilient defense against both current and emerging cyber threats.